\documentclass[%
 aip,
 amsmath,amssymb,
 preprint,%
]{revtex4-1}

\usepackage{graphicx}
\usepackage{dcolumn}
\usepackage{bm}

\usepackage[utf8]{inputenc}
\usepackage[T1]{fontenc}
\usepackage{mathptmx}
\usepackage{color}

\begin{document}

\title{Electric Field Effect on the Thermal Conductivity of Wurtzite GaN}

\author{Yujie Quan}
\author{Sheng-Ying Yue}%
\author{Bolin Liao}%
\email{bliao@ucsb.edu.}
\affiliation{ 
Department of Mechanical Engineering, University of California, Santa Barbara, CA 93106, USA
}%

\date{\today}

\begin{abstract}
Gallium nitride (GaN), a wide band-gap semiconductor, has been broadly used in power electronic devices due to its high electron mobility and high breakdown voltage. Its relatively high thermal conductivity makes GaN a favorable material for such applications, where heat dissipation is a major concern for device efficiency and long-term stability. However, in GaN-based transistors, where the active region can withstand extremely strong electric fields, the field effect on the thermal transport properties has drawn little attention so far. In this work, we apply first-principles methods to investigate phonon properties of wurtzite GaN in the presence of a near-breakdown electric field applied along different crystallographic directions. We find that the electric field changes thermal conductivity considerably via impacting the bond stiffness and ionicity as well as the crystal symmetry, although it has little effect on phonon dispersions. The presence of an out-of-plane electric field increases (decreases) the thermal conductivity parallel (perpendicular) to the electric field, which is attributed to different changes of the Ga-N bond stiffness and ionicity. When an in-plane electric field is applied, the sizable decrease of thermal conductivities along all directions is attributed to the crystal symmetry breaking that enhances the phonon-phonon scattering. Our study provides insights into the effect of extreme external electric fields on phonon transport properties in wide-gap semiconductors.

\end{abstract}

\maketitle

Excellent electronic properties of GaN, such as the wide band-gap and thus the high breakdown electric field, as well as the high electron mobility, make it a promising material for high-power and high-frequency applications \cite{chowdhury2013lateral,amano20182018,fay2020high}. Due to the extremely high power densities present in GaN-based power electronic devices\cite{zhang2013electrothermal}, efficient thermal management is of great importance to maintain the device efficiency and avoid device failure due to large local temperature rises. Extensive experimental and theoretical studies have shown that bulk wurtzite GaN possesses a high thermal conductivity \cite{PhysRevLett.109.095901,garg2018spectral,yang2016nontrivial,slack2002some,shibata2007high,paskov2017effect}, which is beneficial for heat dissipation in power electronic applications. Recently, new strategies to form high-quality bonds between GaN and high-thermal conductivity substrates with low interfacial thermal resistances are also being developed\cite{cheng2020interfacial}. However, these thermal transport studies so far have focused on GaN without an external electric field. Given the extreme electric field strength (up to a few MV/cm)\cite{zhang2013electrothermal,bagnall2016electric} present in the active region in GaN-based devices, the impact of the external electric field on the thermal conductivity of GaN is critical to evaluate the thermal performance of operating devices. 

The electric field effect on thermal conductivity is not well understood in general.
Computationally, recent advances in first-principles methods based on density functional theory (DFT), density functional perturbation theory \cite{baroni2001phonons} (DFPT), and phonon Boltzmann transport equation (BTE) have facilitated the calculation of thermal conductivities of a wide range of crystalline materials under zero electric field\cite{broido2007intrinsic,lindsay2016first}. However, there have been very few studies about the influence of an external electric field on thermal transport properties. Most of these studies examined  two-dimensional (2D) materials in the presence of an out-of-plane electric field. For these studies, slab supercells with a sawtooth potential\cite{kunc1983external} were used to incorporate finite electric fields. This method is natural for simulating 2D materials but becomes computationally expensive for bulk three-dimensional materials due to the large required supercell size. Qin et al.\cite{qin2017external} showed that the thermal conductivity of silicene can be tuned within a wide range, up to two orders of magnitude, with an out-of-plane external electric field that modulates the interactions between atoms. Similar results were also obtained in monolayer borophene by Yang et al.\cite{yang2020electric}. Experimental studies have mostly focused on ferroelectric materials so far\cite{ihlefeld2015room,foley2018voltage}.  

\begin{figure}
\includegraphics[scale=0.7]{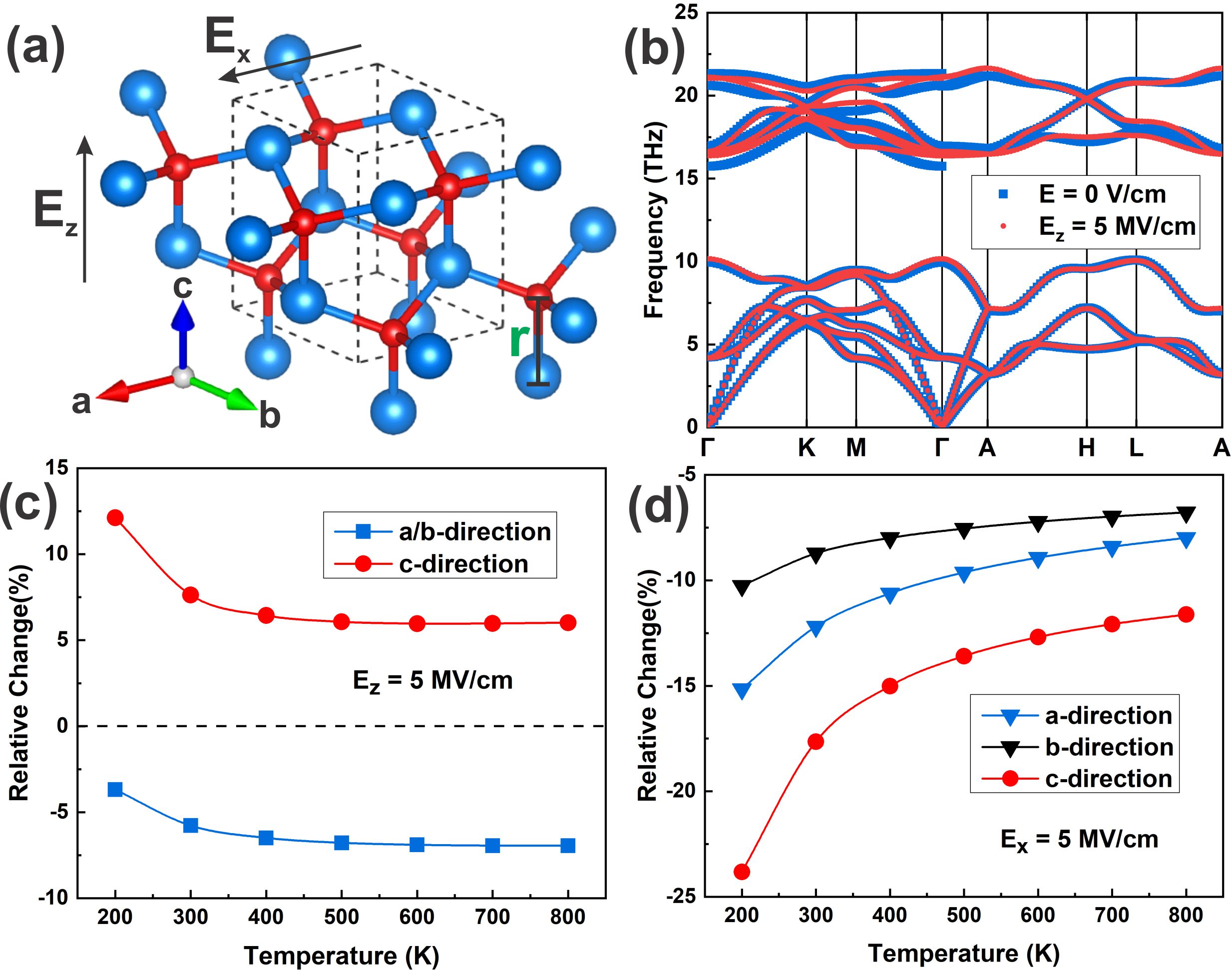}
\caption{\label{fig:fig1} (a) The crystal structure of wurtzite GaN. The applied electric field directions are labeled. (b) The comparison of phonon dispersions of wurtzite GaN with and without an out-of-plane electric field. (c) and (d): The relative changes of the calculated lattice thermal conductivity along different directions in the presence of (c) out-of-plane and (d) in-plane electric fields. }
\end{figure}

The development of the modern theory of polarization\cite{PhysRevLett.89.157602,PhysRevLett.89.117602,wang2006first}, where an electronic enthalpy functional including an additional polarization term is minimized with respect to polarized Bloch functions, has provided a practical first-principles method to calculate the response of physical properties to a finite electric field in 3D bulk materials, since supercells are not required. In this Letter, we apply this first-principles method to investigate phonon properties of wurtzite GaN in the presence of a near-breakdown electric field applied along different directions. The electric field is applied along both the in-plane direction (the $a$-axis) and the out-of-plane direction (the $c$-axis), as shown in Fig.~\ref{fig:fig1}(a). For both cases, we found considerable changes of the thermal conductivity, up to 24\% at 200 K and 13\% at 300 K, depending on the directions of the electric field and the thermal conductivity [Fig.~\ref{fig:fig1}(c) and (c)]. Given the high heat current existing in power electronic devices, this moderate change of thermal conductivity can lead to sizable temperature variation in the device. Through detailed analysis of individual phonon modes, we attributed the change of the thermal conductivity to the combined effect of two major factors: the change of bond ionicity and stiffness and the crystal symmetry breaking due to the applied electric field. The change of bond stiffness and ionicity affects the phonon frequency and group velocity, while the broken crystal symmetry leads to enhanced phonon-phonon scatterings. Our study suggests that the field effect on the thermal conductivity in wide-gap semiconductors needs to be considered in designing thermal management strategies for high-power devices based on these materials.

DFT calculations were performed with the Quantum ESPRESSO (QE) package\cite{giannozzi2009quantum} with the scalar-relativistic Optimized Norm-Conserving Vanderbilt (ONCV) pseudopotentials\cite{PhysRevB.88.085117} within the local-density approximation (LDA). The kinetic energy cutoff for wavefunctions is set to 80 Ry. A mesh grid of $\rm 10\times 10\times 5$ in the first Brillouin zone (BZ) is adopted and the total electron energy convergence threshold for self-consistency is $\rm 1\times 10^{-10}~Ry$. The crystal lattice is fully relaxed with a force-threshold of $\rm 1\times 10^{-4} eV/\textup{\AA}$, with zero-field lattice parameters $a=3.16 \textup{\AA}$, $c=5.148 \textup{\AA}$ and the internal parameter $u=0.375$ ($u$ is the ratio between the Ga-N bond length $r$ along the $c$-axis and the lattice parameter $c$), in excellent agreement with experimental values\cite{schulz1977crystal}. The second- and third-order interatomic force constants (IFCs) were calculated using a finite displacement approach\cite{esfarjani2011heat} with $3\times 3 \times 2$ supercells with or without the electric field. The phonon dispersions based on the second-order IFCs were calculated using the PHONOPY package\cite{PhysRevB.78.134106}, and the non-analytical correction term due to the long-range Coulomb interactions was also included. The interactions between atoms were taken into account up to fifth-nearest neighbors in third-order IFC calculations. The phonon-phonon scattering rates and phonon-isotope scattering rates can be computed based on the IFCs, as detailed in previous works\cite{PhysRevLett.109.095901,garg2018spectral}. We calculated the lattice thermal conductivity by solving the phonon Boltzmann Transport Equation (BTE) iteratively as implemented in ShengBTE\cite{li2014shengbte}. The q-mesh samplings for the BTE calculations were $15 \times15 \times 15$. The convergence of the lattice thermal conductivity with respect to the q-mesh density and the interaction distance cutoff was confirmed. The finite electric field was applied based on the modern theory of polarization\cite{PhysRevLett.89.157602,PhysRevLett.89.117602} as implemented in QE. Given the 3.4 eV band-gap of the wurtzite GaN, the near-breakdown electric field $5$ MV/cm is applied in the calculation. The crystal structure is fully relaxed with the applied electric field. The change of the lattice parameters $a$ and $c$ with the electric field is described by the inverse piezoelectric coefficients of GaN: $d_{31}=\frac{\partial \ln(a)}{\partial E_z}=-1.37 \times 10^{-4}$ cm/MV and $d_{33}=\frac{\partial \ln(c)}{\partial E_z}=2.83 \times 10^{-4}$ cm/MV\cite{bagnall2016electric}. Therefore, the change of the lattice parameters is under 0.1\% with the applied electric field, which agrees with our calculation. It is noted here that the electric field is also applied during the calculations of second- and third-order IFCs, therefore our calculations can capture the full response of phonon properties to the applied electric field, including the effects of both the change of atomic positions and the redistribution of the electronic charge density\cite{wang2006first}. 

The comparison of phonon dispersions with and without applied electric field along the $c$-axis is shown in Fig.~\ref{fig:fig1}(b). A similar plot with the electric field applied along the $a$-axis is given in Fig. S1 (see Supplementary Information). In the presence of the electric field along both $a$-axis and $c$-axis, the dispersions of low-frequency acoustic phonons are not changed appreciably, while slight changes arise for the high frequency optical phonons. Our calculated shift of the optical phonon frequencies at $\Gamma$ point with an out-of-plane electric field agrees qualitatively with the Raman measurements by Bagnall et al.\cite{bagnall2016electric}, where the frequency of A1 (LO) mode increases and that of E2 (high) and E2 (low) modes decrease, validating our calculations incorporating the electric field. 

Although the electric field has little effect on the phonon dispersion, it has a much greater impact on the lattice thermal conductivity, as shown in Fig.~\ref{fig:fig1}(c) and (d). The thermal conductivity of GaN as a function of temperature in zero field is shown in Fig. S2 (see Supplementary Information). Both phonon-phonon scattering and phonon-isotope scattering are considered when calculating the thermal conductivity, since it is well known that the phonon-isotope scattering plays a prominent role in GaN\cite{PhysRevLett.109.095901}. Our calculated thermal conductivity of isotopically pure GaN is roughly 410 W/mK at 300K, while the calculated thermal conductivity of GaN with the natural abundance of Ga isotopes is around 250 W/mK at 300 K, agreeing well with other first-principles calculations\cite{PhysRevLett.109.095901,garg2018spectral,yang2016nontrivial}. When the electric field is applied along the out-of-plane direction, a sizable increase of the thermal conductivity parallel to the electric field is observed, with more than 12\% increase at 200K and around 8\% at higher temperatures. In contrast, the thermal conductivity perpendicular to the electric field decreases by roughly 7\%. On the other hand, when the electric field is applied along the in-plane direction, the in-plane crystal symmetry is broken, which makes $a$-axis and $b$-axis no longer equivalent. Moreover, the thermal conductivities along all three axes decrease in this case, with the largest reduction reaching 24\% at 200 K for the thermal conductivity along the $c$-axis. 

\newcommand{\tabincell}[2]{\begin{tabular}{@{}#1@{}}#2\end{tabular}}  
\begin{table*}
\caption{\label{tab:table3}Relative changes of Ga-N bond properties with and without an out-of-plane near-breakdown electric field of $5$ MV/cm.}
\begin{ruledtabular}
\begin{tabular}{ccccc}
 &\tabincell{c}{In-plane Ga-N\\ bond length}&\tabincell{c}{In-plane Ga-N\\ bond stiffness}&\tabincell{c}{Out-of-plane \\ Ga-N bond length }&\tabincell{c}{Out-of-plane \\ Ga-N bond stiffness }
 \\ \hline
 Relative change &0.3\%&-2.3\%&-0.9\%& 6.4\%\\
 
\end{tabular}
\end{ruledtabular}
\end{table*}

Since the lattice thermal conductivity can be decomposed into contributions from individual phonon modes\cite{esfarjani2011heat}, we analyze the transport and scattering properties of individual phonon modes to understand the mechanisms underlying the observed field effect. We first examine the case where an out-of-plane electric field is applied. Since the field direction is aligned with the rotational axis ($c$-axis) of the crystal lattice, the full crystal symmetry of GaN is preserved. The phonon-phonon scattering rates in this case are presented in Fig. S3 (see Supplementary Information), showing no appreciable difference with or without the applied field. The change of the spectral contribution to the thermal conductivity (Fig. S3) also shows no clear trend for different frequency regimes. This result indicates that both the bond anharmonicity and the phonon scattering selection rules are not significantly affected by the out-of-plane field. Then, we focus on the bond length and the bond strength of the Ga-N bonds, as they can influence the harmonic phonon properties, such as the mode-specific heat capacity and the group velocity. The data is summarized in Table I. We notice a slight decrease of the bond length of the out-of-plane Ga-N bond by 0.9\% and an increase of the bond length of the in-plane Ga-N bond by 0.3\%. The change of the out-of-plane bond length is on the same order as the experimental value\cite{bagnall2016electric}. Using second-order polynomial fitting of the total energy as a function of small perturbations (smaller than $0.05\textup{\AA}$) of the bond length, we obtain the stiffness of the in-plane and out-of-plane Ga-N bonds. We find that the stiffness of the out-of-plane Ga-N bond is increased by 6.4\%, while the stiffness of the in-plane Ga-N bond is decreased by 2.3\%. Typically, the increase of the bond stiffness leads to the enhancement of the phonon frequency and the group velocity. Although the change of phonon frequency is not clear in Fig.~\ref{fig:fig1}(b), where all phonon modes are displayed, the change is more evident for the phonon modes vibrating along a certain bond direction. The phonon modes vibrating along the out-of-plane direction are selected according to their eigenvectors, and the corresponding phonon frequency and group velocity compared with the zero-field values are shown in Fig. S4 (see Supplementary Information).  Considering the negligible contribution of high-frequency optical phonons to the thermal conductivity, only phonons with frequencies below 10 THz are shown. It is found that both the out-of-plane group velocities and the frequency of this group of phonons are increased by 2\% on average in the presence of the out-of-plane electric field, which are the main contributors to the increased out-of-plane thermal conductivity. In contrast, the frequencies of the phonons with in-plane vibrations are reduced. Although the in-plane group velocities of some of the phonons with in-plane vibrations are slightly increased with the electric field, their frequency lies in the range of 5-7 THz and they are known to be strongly suppressed by phonon-isotope scatterings\cite{garg2018spectral}. 

\begin{figure}
\includegraphics[scale=0.3]{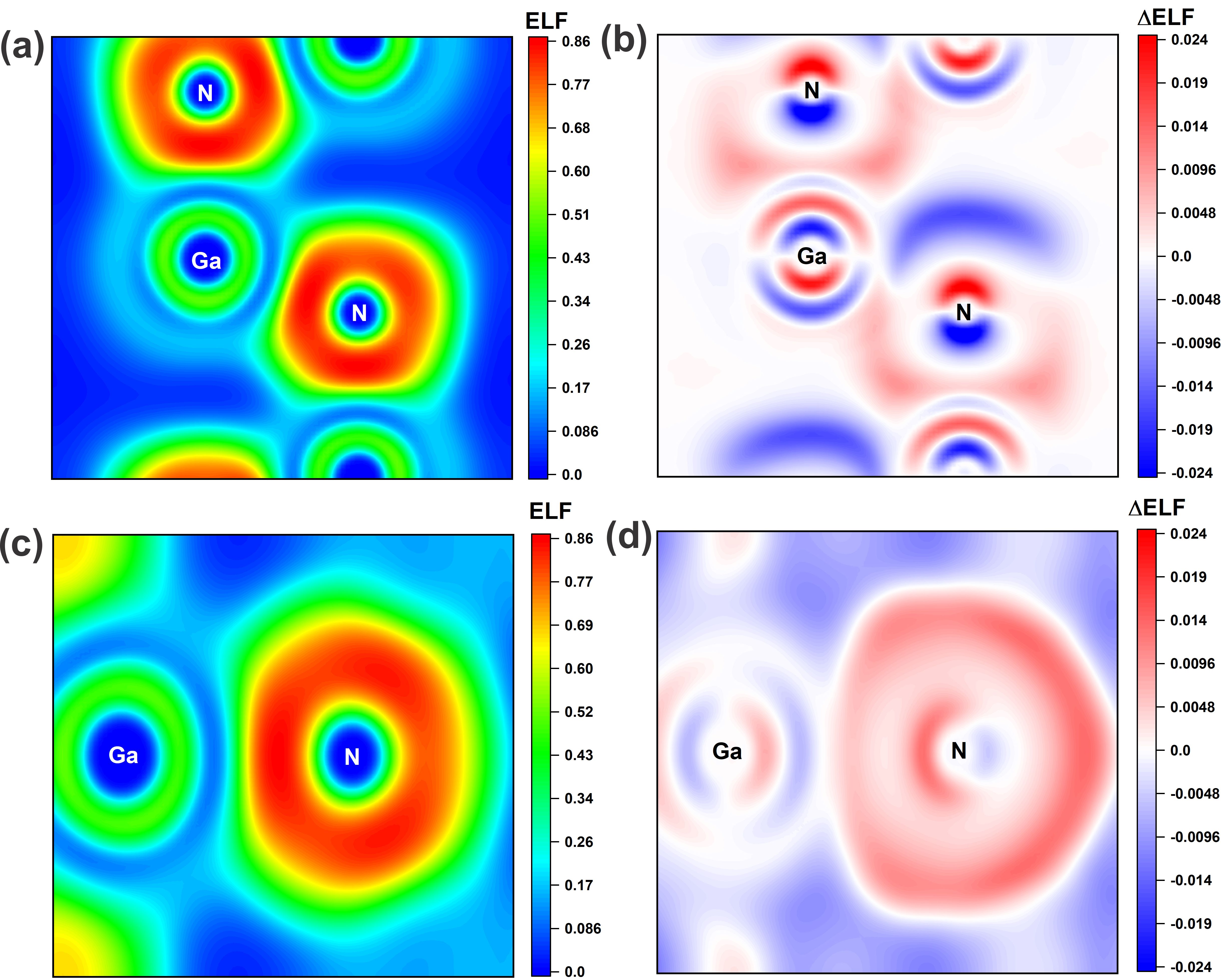}
\caption{\label{fig:fig-2} The electron localization function (ELF) at zero field and its change in the presence of the out-of-plane electric field. (a) and (b) a cross section along the (110) surface showing the ELF around an out-of-plane Ga-N bonds. (c) and (d) a cross section determined by the in-plane Ga-N bond direction and the [100] crystallographic direction, showing the ELF around an in-plane Ga-N bond. The ELF values are coded in the color bars.}
\end{figure}

Given that the relative change of the bond stiffness is much greater than that of the bond length, we hypothesize that the change of the bond stiffness in the presence of the electric field is not only a consequence of the change of the bond length, but also affected by the redistribution of the charge density, i.e. the ionicity of the Ga-N bonds. The electron localization function (ELF) is often used as a visually informative method to demonstrate the bond ionicity\cite{savin1997elf}, whose value is normalized between 0 and 1. A higher EFL value at a spatial location indicates a stronger Pauli exclusion to a test electron placed in the proximity, signaling a higher local charge density and bond saturation. Therefore, higher EFL values concentrated in between two ions indicate strong covalent bonding, while low EFL values between the two ions are a signature of highly ionic bonds\cite{savin1997elf}. The ELF at zero field and its change in the presence of the out-of-plane electric field are shown in Fig. \ref{fig:fig-2}. Fig. \ref{fig:fig-2}(a) and (b) are ELF describing the out-of-plane Ga-N bonds and its change due to the electric field, while Fig. \ref{fig:fig-2}(c) and (d) provide the same information for the in-plane Ga-N bonds. The Ga-N bonds show a strong ionic characteristic, where very low electron density is found in the middle of the bond, implying very few shared electrons, as shown in Fig. \ref{fig:fig-2}(a) and (c). The applied external electric field modifies the distribution of electrons and results in an increase of ELF in the middle of the out-of-plane Ga-N bond and a decrease near the nuclei, as shown in Fig. \ref{fig:fig-2}(b), implying that the charge density in the bonding region increases and thus the ionicity of this out-of-plane Ga-N bond is decreased. In contrast, for the in-plane Ga-N bonds, the ionicity increases with the electric field, which is reflected in the increased charge density near the N atom. It is understood that more ionic bonds usually possess lower bond stiffness and thus lead to a lower lattice thermal conductivity\cite{joffe1956heat}, and thus the impact of the applied electric field on the bond ionicity in GaN also contributes to the change of the thermal conductivity.

So far, we have analyzed the change of the bond stiffness and ionicity in the presence of the out-of-plane electric field. We conclude that the out-of-plane electric field, under which the crystal symmetry is preserved, increases the thermal conductivity along the out-of-plane direction via enhancing the strength of the Ga-N bonds along the same direction and reducing its ionicity. In the mean time, the out-of-plane electric field decreases the thermal conductivity perpendicular to the electric field through the opposite effects.

\begin{figure}
\includegraphics[scale=0.8]{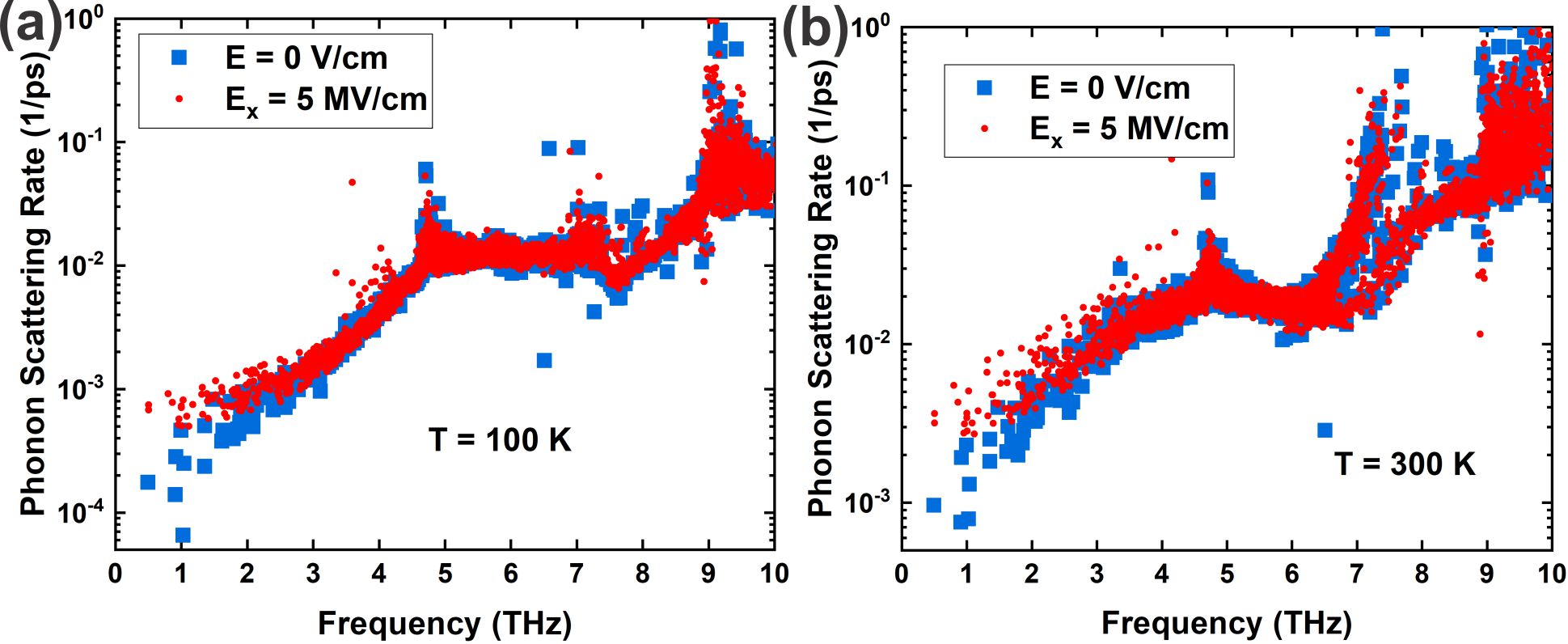}
\caption{\label{fig:fig-3} Calculated phonon-phonon scattering rates at 100K and 300 K with and without the applied in-plane electric field of $5$ MV/cm. The low-frequency acoustic phonons are mostly affected by the symmetry-breaking electric field.}
\end{figure}

When the electric field is applied along the in-plane direction, the thermal conductivities along all directions show a significant reduction. Different from the out-of-plane electric field, under which the crystal symmetry is preserved, the in-plane electric field breaks the crystal symmetry. The space group of wurtzite GaN crystal structure is $P6_3mc$, while after the full lattice relaxation with the in-plane electric field of $5$ MV/cm, the space group of the crystal structure becomes $P1$, indicating the complete absence of any rotation axes, rotary-inversion axes, screw axes, or mirror planes. Due to the symmetry breaking, selection rules for phonon-phonon scatterings imposed by the crystal symmetry are lifted, contributing to more allowed phonon-phonon scattering channels. A recent study found a 20\% reduction of the thermal conductivity of GaAs under a small symmetry-breaking strain\cite{vega2019reduced}. Since the zone-center phonons possess the same symmetry operations as the full point group of the crystal, low-frequency acoustic phonons near the zone center are expected to be affected more strongly. To verify this effect in GaN with the in-plane electric field, we calculate the phonon-phonon scattering rates with or without the field at 100 K and 300 K, which are shown in Fig. \ref{fig:fig-3}. The scattering rates of low frequency acoustic phonons in the presence of the in-plane electric field are enhanced compared to the zero-field scattering rates, which contributes to the reduction of the thermal conductivities along all three directions. This effect is more prominent at lower temperatures (we are not reporting the thermal conductivity at 100 K due to the computational cost to obtain a converged thermal conductivity at lower temperatures).  

In summary, we demonstrated the impact of a near-breakdown electric field on phonon transport properties of wurtzite GaN through first-principles calculations. We find that the electric field changes the thermal conductivity considerably by affecting the bond stiffness and ionicity as well as the crystal symmetry, depending on the directions of the applied electric field and the thermal conductivity.  Our study points out the importance of the field effect on the thermal conductivity of wide-gap semiconductors and contributes to the future development of electrical and thermal co-design strategies of next-generation power electronic devices\cite{tsao2018ultrawide}.

The data that support the findings of this study are available from the corresponding author upon reasonable request.

This work is based on research supported by the National Science Foundation under the award number CBET-1846927. Y. Q. acknowledges the support from the Graduate Traineeship Program of the UCSB NSF Quantum Foundry via the Q-AMASE-i program under award number DMR-1906325. This work used the Extreme Science and Engineering Discovery Environment (XSEDE)\cite{towns2014xsede}, which is supported by National Science Foundation grant number ACI-1548562.

\nocite{*}
\bibliography{aipsamp}

\appendix
\renewcommand{\thefigure}{S\arabic{figure}}
\setcounter{figure}{0}
\newpage
\clearpage
\begin{figure}
\includegraphics[scale=1]{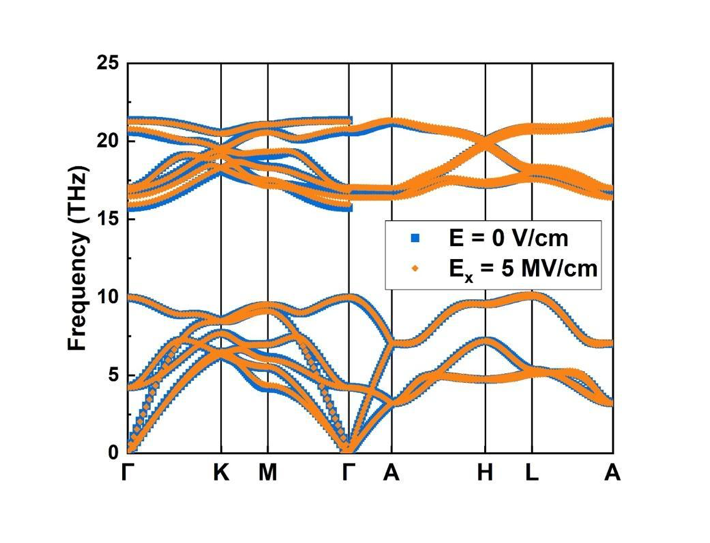}
\caption{\label{fig:fig-s1} The phonon dispersions of GaN with and without the applied electric field along the in-plane direction.}
\end{figure}

\newpage
\clearpage
\begin{figure}
\includegraphics[scale=1]{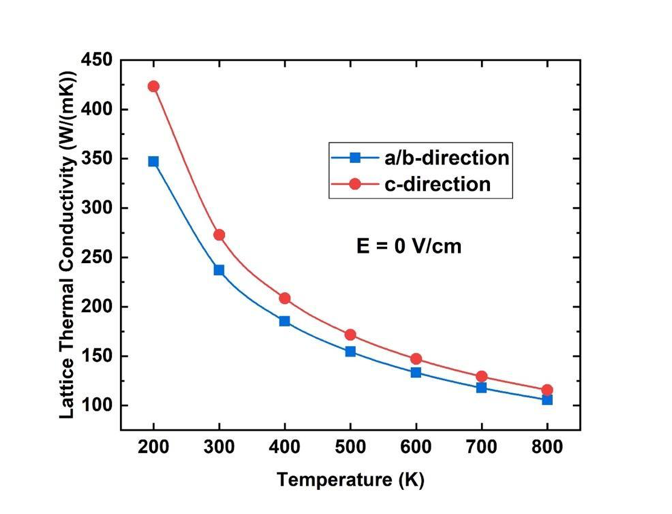}
\caption{\label{fig:fig-s2} The calculated lattice thermal conductivity of wurtzite GaN at zero field, including both phonon-phonon and phonon-isotope scatterings (with the natural abundance of Ga isotopes in GaN).}
\end{figure}

\newpage
\clearpage
\begin{figure}
\includegraphics[scale=1]{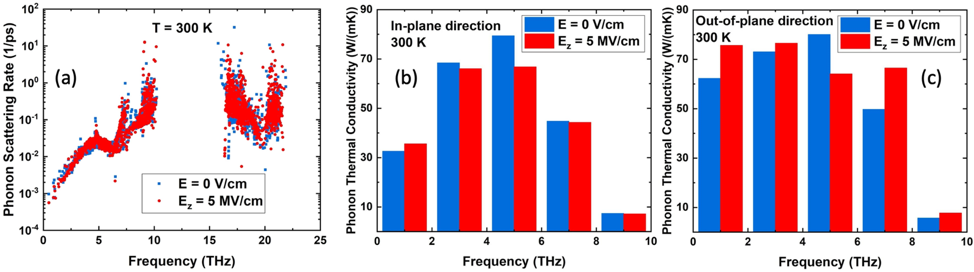}
\caption{\label{fig:fig-s3} (a) The phonon-phonon scattering rate at 300K at zero field (blue spots) and with an out-of-plane 5 MV/cm electric field. (b) The change of the spectral contribution to the in-plane thermal conductivity with an out-of-plane 5 MV/cm electric field. (c) The change of the spectral contribution to the out-of-plane thermal conductivity with an out-of-plane 5 MV/cm electric field.}
\end{figure}

\newpage
\clearpage
\begin{figure}
\includegraphics[scale=0.6]{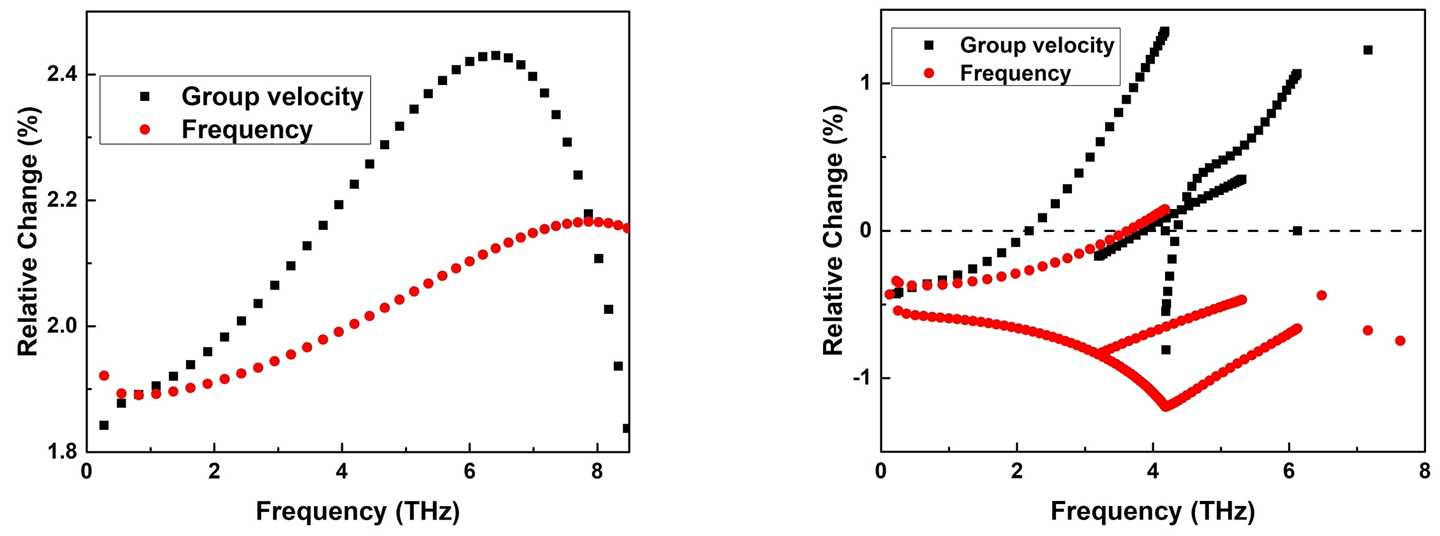}
\caption{\label{fig:fig-s4} (left) The relative change of the out-of-plane group velocity and frequency of the phonons vibrating along out-of-plane direction in the presence of an out-of-plane electric field. (right) The relative change of the in-plane group velocity and frequency of the phonons vibrating along in-plane direction in the presence of an out-of-plane electric field.}
\end{figure}

\end{document}